\documentclass[12pt,draftcls,onecolumn]{IEEEtran}
\ifCLASSINFOpdf
\else
\fi
\usepackage{amssymb}
\usepackage{amsmath}
\usepackage{enumerate}
\usepackage{graphicx}
\usepackage{epsfig}
\usepackage{multirow}
\usepackage{algorithm}
\usepackage{algorithmic}
\usepackage{setspace}
\usepackage{booktabs}
\usepackage{bm}
\usepackage{color}
\usepackage{flushend}
\usepackage[footnotesize]{caption2}
\usepackage{fancyhdr}

\begin{document}

\title{Green Resource Allocation and Energy Management in Heterogeneous Small Cell Networks
Powered by Hybrid Energy}
\author{Qiaoni Han, Bo Yang, Nan Song, Yuwei Li， and Ping Wei
\thanks{Q. Han is with the Department of Automation, Tianjin University, Tianjin 300072, China; B. Yang, N. Song, Y. Li, and P. Wei are with the Key Laboratory of System Control and Information Processing, Ministry of Education of China, and the Department of Automation, Shanghai Jiao Tong University, Shanghai 200240, China.}}
%\thanks{Corresponding author: Bo Yang.}
\maketitle
\thispagestyle{fancy}
\fancyhead[L]{DOI: 10.1016/j.comcom.2020.06.002}
\cfoot{}
\vspace{-0.5cm}
\begin{abstract}
In heterogeneous networks (HetNets), how to improve spectrum efficiency is a crucial issue. Meanwhile increased energy consumption inspires network operators to deploy renewable energy sources as assistance to traditional electricity. Based on above aspects, we allow base stations (BSs) to share their licensed spectrum resource with each other and adjust transmission power to adapt to the renewable energy level. Considering the sharing fairness among BSs, we formulate a multi-person bargaining problem as a stochastic optimization problem. We divide the optimization problem into three parts: data rate control, resource allocation and energy management. An online dynamic control algorithm is proposed to control admission rate and resource allocation to maximize the transmission and sharing profits with the least grid energy consumption. Simulation results investigate the time-varying data control and energy management of BSs and demonstrate the effectiveness of the proposed scheme. \newline
\end{abstract}

\vspace{-0.1cm}

\section{Introduction}

The explosive growth of global mobile data traffic and mounting energy
problems have led to higher requirement for both area spectral efficiency
and energy efficiency\cite{1}. Future cellular networks are now rapidly
evolving to heterogeneous network (HetNet) architectures with
small-cell base stations (SBSs) improving quality of service (QoS) and
macrocell base stations (MBSs) guaranteeing global coverage area\cite{2}. There are two kinds of spectrum sharing in HetNets: orthogonal sharing and non-orthogonal sharing\cite{3,4,5}. Orthogonal sharing allows BSs to operate in the orthogonal subchannels, which lowers the interference with each other. Non-orthogonal spectrum sharing allows BSs to reuse the available spectrum resources with higher interference at the receivers\cite{6}.

Although the energy consumption of individual SBS is small, a large number of SBSs lead to higher energy consumption. In fact, the communication networks expend approximately 60 billion kWh per year\cite{7}. High energy consumption of HetNets is an urgent problem to be solved at present, so we want to reduce energy consumption without sacrificing network performance. Energy harvesting technology is an effective method to reduce energy cost\cite{8}. However, due to the fluctuation of energy harvesting, the renewable energy generation may not adapt to the load traffic conditions. To guarantee steady communication, the power grid provides auxiliary power supply.

Spectrum is another vital resource in wireless communications, besides energy\cite{9}. However, existing spectrum allocation schemes often lead to a low spectrum utility, because the BSs of different mobile network operators (MNOs) only consider their own spectrum requirements instead of collective interests\cite{10}. Due to the BSs' self-interest, we need to design incentives to encourage BSs with idle spectrum resource to share it to improve spectrum utilization.

Therefore, the wireless resource allocation scheme should be adaptive to the finite spectrum resources and stochastic energy sources with the least grid energy consumption to ensure the QoS of mobile users.

\subsection{Prior Work and Motivation}

Previous researches have provided an overview on resource allocation and interference management in a two-tier HetNet in recent years. Palanisamy and Nirmala \cite{11} have presented a comprehensive survey on downlink interference management strategies in HetNets. Zhang \textit{et al}. \cite{12} studied resource optimization for the interference management and self-organization schemes in a hybrid self-organized small cell network. In \cite{13}, the authors considered the sum-rate optimization problem with power control for uplink transmission in a HetNet, then a new practical near-optimal distributed algorithm that eliminates these network overheads was proposed. Lang \textit{et al}. \cite{14}  adopted the heterogeneous genetic algorithm to solve the resource allocation problem for cognitive decode-and-forward relay network. A game-theoretical scheme using energy-efficient resource allocation and interference pricing was proposed in \cite{15} for an interference-limited environment without complete channel state information.

Apart from designing efficient resource allocation policies, there are some
energy-saving studies for HetNet. The deployment of renewable
energy has been considered in \cite{16,17,18,19}. Gong \textit{et al}. \cite{16}  considered energy-efficient wireless resource management in cellular networks powered by renewable energy. The goal is to minimize the average grid power consumption while satisfying the users' QoS (blocking probability) requirements. It uses statistical information for traffic intensity and renewable energy to adaptively switch BSs' on-off state and adjust the allocated resource blocks. Zhang \textit{et al}. \cite{17} considered coordinated multi-point (CoMP) communication systems and integrated renewable energy sources (RES) into smart grid. A convex optimization problem is formulated to minimize the worst-case energy transaction cost while guaranteeing the QoS of users. Yang \textit{et al}. \cite{18} jointly considered the resource allocation and energy management of HetNet powered by hybrid energy with the consideration of spatio-temporal diversity of traffic and renewable energy. Qin \textit{et al}. \cite{19} investigated the resource allocation for orthogonal frequency-division multiple access (OFDMA)
wireless networks powered by renewable energy and traditional grid. There is
a trade-off between network throughput and grid energy consumption. The
proposed main solution is similar to \cite{18}.

Due to the finite spectrum resources, several spectrum sharing algorithms have been proposed in literature such as \cite{20,21,22,23,24,25}. Specifically, spectrum sharing in the same operator is considered in \cite{20,21,22}. Zhang \textit{et al}. \cite{20} proposed a distributed resource management scheme in a HetNet, which divides the problem into two sub-problems: BS cluster and subchannel allocation. Note that only the BSs in the same cluster are allowed to share the spectrum band. Lee \textit{et al}. \cite{21} utilized the cognitive radio (CR) technique to improve the performance of wireless powered communication network. Two coexisting models for spectrum sharing were proposed and the authors considered the interference between two models to protect the primary user transmission. The coexisting network including device-to-device links and cellular links was adopted and a new game model called Bayesian overlapping coalition formation game was proposed in \cite{22}. Moreover, multi-MNO coordinate their spectrum resources sharing in \cite{23,24,25}. A common spectrum pool to share spectrum was investigated in \cite{23}. It formulates the sharing between the MNOs as a repeated game and determines rules to decide whether to participate in the game at each stage game.
An underlay/overlay (hybrid) transmission mode to share the spectrum of licensed users was introduced in \cite{24}. The objective is to maximize network throughput and eliminate interference. Zhang \textit{et al}. \cite{25} considered the spectrum sharing among multi-MNO in the unlicensed spectrum. A hierarchical game including Kalai-Smorodinsky bargaining game and Stackelberg game is proposed to reduce the interference.

Motivated by the above works, we can use renewable energy to reduce energy consumption and share spectrum to improve spectrum utilization. There are some existing works in this field. But most proposed solutions ignore the
fairness between BSs and time-varying of network states. Although some
studies, e.g. \cite{26} consider a fair resource allocation, they focused
mainly on the network throughput without sharing costs. However, selfish BSs
emphasize on improving their own utility without considering other BSs of different MNOs.
Thus, we need to design incentive mechanism to promote the BSs to share idle spectrum resource.

In this paper, we consider joint subchannel allocation and power control under cross-tier interference, fairness between SBSs and renewable energy constraints. The main contributions of this paper are as follows:

\begin{itemize}
\item First, we jointly consider subchannel allocation and power control
strategy for spectrum sharing in a HetNet, while considering the cross-tier interference constraint.
\item Second, a multi-person bargaining problem is modelled to measure the
fairness among SBSs. An online dynamic resource allocation scheme consisting
of spectrum pricing, admission flow control and resource allocation is
developed by solving a stochastic optimization problem.
\item Third, simulation results are presented to show the effectiveness of the
proposed approach. Our approach can substantially achieve the trade-off
between transmission profits and network throughput.
\end{itemize}

The rest of this paper is organized as follows. In Section II, we illustrate
the system model and constraint conditions. Section III formulates a
stochastic optimization problem based on the Nash bargaining. We present an online dynamic resource allocation
and energy management scheme in section IV. Numerical results are given in Section V with
discussions. Conclusion and future work are presented in Section VI.

\section{System and Queue Model}

\subsection{Network Model}

\begin{figure}[htbp]
\centering
\includegraphics[width=11cm]{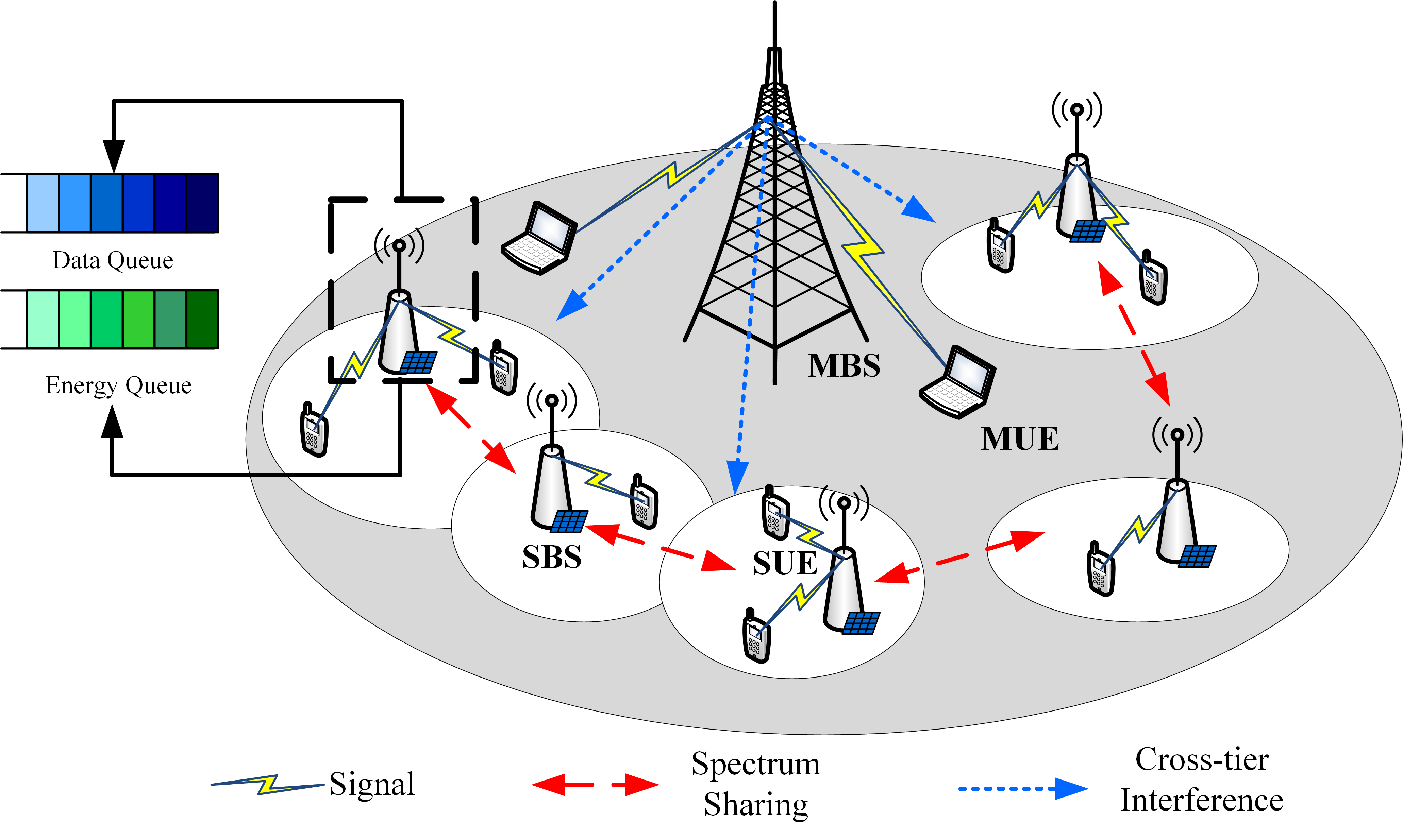}
\caption{An example of the considered network model}
\label{sysmod}
\end{figure}

We consider a downlink OFDMA two-tier HetNet. As shown in Fig.\ref{sysmod},
SBSs belonging to different MNOs are overlaid in the coverage of a MBS. There are two types of
mobile user equipments (UEs): macro-cell UE (MUE) and small cell UE
(SUE) corresponding to the service BSs. All SBSs are assumed to be close
access mode that allows only the authorized SUEs to access the corresponding
SBS. We assume that the SBSs occupy different spectrum bands. Thus, the
co-tier interference between SBSs can be considered to be negligible.

Accordingly, let $\mathcal{N}_0=\left \{ 0,1,2,...,N \right \}$ denote the
BS set, where $n=0$ represents MBS and $\mathcal{N}=\left \{1,2,...,N \right \}$
denotes SBSs. As a consequence, the set of MUEs served by MBS is $\mathcal{U}_{0}$
and the set of SUEs associated with SBS $n$ is $\mathcal{U}_{n}=\left%
\{1,2,3,...,U_{n}\right\}$. The OFDMA system has a bandwidth $F$ divided into
$M$ subchannels, i.e., $\mathcal{M}=\left \{ 1,2,...,M \right \}$ and the bandwidth of subchannel $m
$ is $\varpi_m$, while the spectrum band occupied by SBS $n$ is denoted by $\mathcal{M}_n\in\mathcal{M}$, and $\cup_{n\in\mathcal{N}}\mathcal{M}_n=\mathcal{M}, \mathcal{M}_i\cap\mathcal{M}_j=\varnothing, \forall i\neq j$.
Moreover, since MBS has the priority to choose subchannels, it is assumed that
each subchannel is always occupied by one MUE at a given time, and to guarantee the QoS of MUEs,
the SBSs adjust their power to reduce the cross-tier interference.

As the subchannel conditions and energy harvesting process change dynamically, we consider the HetNet operates in a time-slotted manner where the subchannel and energy harvesting conditions keep stable in a time slot. We use $\mathcal{T}=\left
\{0,1,2,...,t,...,T \right \}$ to denote the time-slot set and the length of
each time slot defaults to 1.

At the beginning of time slot $t$, each SBS observes the channel state
information (CSI) and current queue state information (QSI). We use $%
p_{mk}^M(t)$ to denote the power allocated to MUE $k$ served by MBS on
subchannel $m\in\mathcal{M}_n$ and $h_{nmu}^M(t)$ to denote the interference subchannel gain
from MBS to SUE $u$ in SBS $n$ on subchannel $m\in\mathcal{M}_n$. The received
signal-to-interference-plus-noise ratio (SINR) of the $u$th SUE from SBS $n$
on subchannel $m\in\mathcal{M}_n$ at the $t$-th time slot is given by:
\begin{equation}
\mathit{\gamma}_{nmu}(t)=\frac{p_{nmu}(t)h_{nmu}(t)}{%
p_{mk}^M(t)h_{nmu}^M(t)+\sigma ^{2}}, \forall m\in\mathcal{M}_n
\end{equation}
where $p_{nmu}(t)$ represents the transmission power of SBS $n$ to SUE $u$
on subchannel $m\in\mathcal{M}_n$. $h_{nmu}(t)$ is the channel gain between the SUE $u$ and
SBS $n$ on subchannel $m\in\mathcal{M}_n$ including pathloss, shadowing and other factors. $%
\sigma ^{2}$ denotes the noise power level. Compared with the cross-tier
interference, the co-tier interference can be considered to be de minimis.
Therefore, the SBSs are primarily interfered by the MBS.

The transmission rate of SBS $n$ to SUE $u$ on subchannel $m\in\mathcal{M}_n$ in the
non-cooperative case is
\begin{equation}
R_{nmu}(t)=\varpi_m\mathrm{\log_{2}}\left ( 1+\mathit{\gamma}_{nmu}(t)
\right ), \forall m\in\mathcal{M}_n
\end{equation}

%We assume UEs $k$ is serviced by $BS(n)$.Let binary vatiable $l_{ku}$ denote subchannel allocation state where $l_{ku}=1$ indicates UE $u$ is scheduled on subchannel $k$ and so is $l_{ku}=0$.

Then, the total throughput of SUE $u$ serviced by SBS $n$ can be computed by:
\begin{equation}
R_{nu}(t)=\sum_{m\in\mathcal{M}_n}x_{nmu}(t)R_{nmu}(t)
\end{equation}

We define a binary variable $x_{nmu}(t)$ as subchannel allocation index. $%
x_{nmu}(t)=1$ indicates the subchannel $m\in\mathcal{M}_n$ is allocated to SUE $u$ of SBS $n$
and otherwise, $x_{nmu}(t)=0$. In this transmission model, each subchannel
can be occupied by at most one SUE at the given time. Limited by the
transmission characteristics, we have the following limitations:

\textit{Subchannel allocation constraint}: Each subchannel can be allocated
to at most one SUE in each SBS at time slot $t$ to avoid intra-cell
interference. Each SBS uses different subchannels to avoid inter-tier
interference. Thus, we have
\begin{equation}  \label{equ:sac1}
\sum\limits_{n=1}^{N}\sum\limits_{u=1}^{U_{n}}x_{nmu}(t)\leq 1,\forall m,t
\end{equation}
\begin{equation}  \label{equ:sac2}
x_{nmu}(t)\in \left \{ 0,1 \right \},\forall n,m,u,t
\end{equation}

\textit{SBS power constraint}: Each SBS has a peak transmission power $p_n^{max}$
to avoid the overload. Each SBS's transmission power is limited by
\begin{equation}
\sum\limits_{m\in\mathcal{M}_n}\sum\limits_{u\in\mathcal{U}_n}x_{nmu}(t)p_{nmu}(t)\leq
p_n^{max},\forall n,t
\end{equation}

\textit{Cross-tier interference constraint}: The interference from a SBS to
a MUE occurs when the SBS transmits in the same subchannel occupied by the MUE.
To protect the QoS of MUEs, we denote $I_{m}^{S}(t)$ as the maximum
tolerable cross-tier interference temperature in subchannel $m$ for the assigned MUE $b$
at time slot $t$. Thus, we have
\begin{equation}
\sum\limits_{n=1}^{N}\sum\limits_{u=1}^{U_{n}}x_{nmu}(t)p_{nmu}(t)h_{nmb}(t)%
\leq I_{m}^{S}(t),\forall m,t
\end{equation}
where $h_{nmb}(t)$ is the interference channel gain on subchannel
$m$ from SBS $n$ to MUE $b$ served by the MBS at time slot $t$. The
interference constraint means that SBSs are allowed to transmit signals on the same subchannel with the MBS only if the total interference is kept under a tolerable level.

In this paper, we focus on the joint power control and subchannel allocation
problem to improve network performance.
%\vspace{-0.3 cm}

\subsection{SUE Traffic and Data Queue Model}

We assume there is a data buffer for each SUE to store the arriving packets.
Let the stochastic process $A_{nu}(t)\in \left [ 0,A^{max} \right ]$
denote the arrival traffic amount for SUE $u$ of SBS $n$ in time slot $t$. $%
A_{nu}(t)$ is independent and identically distributed (i.i.d) over different
time slots with the average arrival rate $\lambda_{nu}$, i.e., $\mathbb{E}%
[A_{nu}(t)]=\lambda _{nu}$, where $\mathbb{E}[\cdot ]$ represents the
expectation. To achieve the normal transmission, the admitted traffic amount
$D_{nu}(t)$ has the following constraint:
\begin{equation}
0 \leq D_{nu}(t)\leq A_{nu}(t) \leq A^{max}, \forall n,u,t
\end{equation}

Thus, the finite data queue backlog $Q_{nu}(t)$ of SUE $u$ in small cell $n$
is formulated as
\begin{equation}
Q_{nu}(t+1)=\left [ Q_{nu}(t)-R_{nu}(t) \right ]^++D_{nu}(t)
\end{equation}
where $R_{nu}(t)$ and $D_{nu}(t)$ are the output and input rate of data
queue $Q_{nu}$ respectively and $\left [ x \right ]^{+}=max\left \{x,0
\right \}$. At the beginning, we assume $Q_{nu}(0)=0$. Due to the
time-varying subchannels and packet arrival process, the arrival and departure
processes are both stochastic. Thus, the $Q_{nu}$ is varying over time slot. Let $R(t)$ and $D(t)$ be the output rate and input rate at time slot t for queue $Q(t)$, respectively. For discrete time process $Q(t)$ evolves as following,
\begin{equation}
Q(t+1)=\left [ Q(t)-R(t) \right ]^++D(t)
\end{equation}
we need to guarantee the queue stability. According to \cite{27}, a queue is
defined as strongly stable if
\begin{equation}
\underset{T\rightarrow \infty }{\lim}\frac{1}{T}\sum_{t=0}^{T-1}\mathbb{E}%
\left [ Q(t) \right ]< \infty
\end{equation}

A network is called strongly stable if all the individual queues are
strongly stable. Therefore, it is a crucial prerequisite for the network to
guarantee the stability of all queues.

\subsection{Energy Harvesting and Energy Queue Model}

According to the EARTH project,we adopt a linear approximation power
consumption as\cite{28}:
\begin{equation}
P_{n}(t)=P_{n}^{c}+\Delta_{n}\sum\limits_{u=1}^{U_{n}}%
\sum_{m=1}^{M}x_{nmu}(t)p _{nmu}(t)
\end{equation}
where $P_{n}^{c}$ is the static power consumption including the cooling
system, baseband processor and so on. $\Delta_{n}$ is the slope of the
transmission power dependent power consumptions. According to the maximum
power constraint (6), the power consumption satisfies the following
constraint:
\begin{equation}
P_{n}(t)\leq P_{n}^{c}+\Delta_{n}p_n^{max}\overset{\Delta }{=}P_n^{max}
\end{equation}

To avoid the transmission interruption, the MBS uses a conventional grid
source and the SBSs are powered by not only renewable energy generation but
also the power grid. Each SBS harvests energy from ambient energy sources
such as solar and stores the energy in its battery. If the storage energy is
not enough for transmission, the SBS will be supplied by the grid. We model
the energy harvest process as a random process. Let $E_{n}(t)$ denote the harvested energy in time slot $t$ in small cell $n$. $E_{n}(t)$ is assumed to be i.i.d. with the maximum value $E_n^{max}$.
Notice that SBS may have no a priori knowledge of energy harvest
process, which is appropriate for the situation.

At time slot $t$, the battery level $S_{n}(t)$ is
determined by previous battery level, harvested energy and SBS power
consumption as follows:
\begin{equation}
S_{n}(t+1)=\left[ S_{n}(t)-F_{n}(t)\right]^+ + J_{n}(t)
\end{equation}
where $F_n(t)$ and $J_n(t)$ are the discharge energy and charge energy of
the battery device in SBS $n$ at time slot $t$ respectively. The
storage energy cannot exceed the capacity of the battery, i.e., $S_n(t)\leq
S_n^{max}$.

The power consumption of SBS $n$ consists of two parts: discharge energy from
battery devices $F_n(t)$ and power grid $G_n(t)$. Then, we have
\begin{equation}
P_n(t)=F_n(t)+G_n(t)
\end{equation}

The discharge energy should be less than the battery energy level and the charge energy should consider the battery capacity. Thus, the discharge energy and charge energy should satisfy the following constraints:
\begin{equation}
0 \le {F_n}\left( t \right) \le {S_n}\left( t \right)
\end{equation}
\begin{equation}
0\leq J_n(t)\leq \min\left \{ S_n^{max}-S_n(t),E_n(t) \right \}
\end{equation}

Let $\overline{F}_{n}=\underset{T\rightarrow \infty }{\lim} \frac{1}{T}%
\sum_{t=0}^{T-1}\mathbb{E}\left \{ F_{n}(t) \right \}$ and $\overline{J}_{n}=%
\underset{T\rightarrow \infty }{\lim} \frac{1}{T}\sum_{t=0}^{T-1}\mathbb{E}%
\left \{ J_{n}(t) \right \}$ denote the time-average discharge energy and
the time-average charge energy respectively. The stability of energy queue $S_n(t)$ should be guaranteed.

\subsection{Spectrum Sharing}
At each time slot, the common spectrum pool formed by MNOs from reusing the spectrum of MBS is partitioned between the SBSs. The total spectrum band allocated to SBS $n$ at time slot $t$ is $%
B_{n}(t)=\sum\limits_{u\in\mathcal{U}_n}\sum\limits_{m\in\mathcal{M}_n}\varpi_m x_{nmu}(t)$.
We assume that each SBS uses the same amount of subchannels without any payment in the
initial state. However, the SBS pays for an extra payment if it uses more subchannels than it has put in the common pool. Similarly, a SBS can rent its free spectrum resource to other SBSs for benefits.

We define a utility of SUE $u$ of SBS $n$ in terms of admitted data
rate to represent the satisfaction that SBS $n$ sends data to SUE $u$, then the \textit{time-average expected profit} of SBS $n$ is given as
\begin{equation}
\overline{C}_{n}=\sum_{u=1}^{U_{n}}\overline{D}_{nu}(t)+\overline{O}%
_{n}-\phi \overline{G}_{n}
\end{equation}
where
\begin{equation}
\overline{G}_{n}=\underset{T\rightarrow \infty }{\lim}\frac{1}{T}%
\sum_{t=0}^{T-1}\mathbb{E}\left\{G_n(t)\right\} \\
\end{equation}
\begin{equation}
\overline{O}_{n}=\underset{T\rightarrow \infty }{\lim}\frac{1}{T}%
\sum_{t=0}^{T-1}\mathbb{E}\left\{O_n(t)\right\} \\
\end{equation}
\begin{equation}
O_n(t)=\alpha _{n}(t)\left ( B_{n}^0-B_{n}(t) \right )^+-\beta _{n} (t)\left
( B_{n}(t)-B_{n}^0 \right )^+
\end{equation}
$\phi$ is a constant to
balance gains from spectrum capacity and power consumption, which is a constant.
$B_{n}^0$ denotes the initial spectrum band of SBS $n$. $%
\alpha_{n}(t)$ is the per-unit price of spectrum transferred from SBS $n$ to other SBSs and $\beta_{n}(t)$ is the per-unit price of extra spectrum rented from other SBSs. The spectrum prices all have the maximum $q_n^{max}$. Notice that during any time slot, the SBS $n$ can only choose one spectrum sharing scheme: lease or rent spectrum band.

\section{Problem Formulation}
In this paper, our objective is to maximize the total time-average profit gains
of SBSs (or MNOs) by sharing their spectrum resources with each other and minimizing the grid power consumption at the same time. Furthermore, we need to allocate the spectrum resources fairly so that the SBSs have an incentive to share their spectrum.
Meanwhile the proposed resource allocation and energy management scheme should satisfy interference
constraints, energy constraints and network stability. Then, based on the Nash bargaining problem studied in \cite{29},
the optimization problem can be formulated as
\begin{equation}
\begin{split}
\mathrm{OP}_1: &\quad\max\limits_{\mathbf{D,X,P,G},\boldsymbol{\alpha,\beta}}\sum\limits_{n=1}^{N}\log(%
\overline{C}_{n}-{C_{n}^{min}})\\
\text{s.t.}\quad &C1: \overline{C}_{n}\geq C_{n}^{min},\forall n \\
&C2:0 \leq D_{nu}(t)\leq A_{nu}(t) \leq A^{max}, \forall n,u,t\\
&C3:\sum\limits_{m=1}^{M}\sum\limits_{u=1}^{U_{n}}x_{nmu}(t)p_{nmu}(t)\leq
p_n^{max} ,\forall n \\
&C4:\sum\limits_{n=1}^{N}\sum%
\limits_{u=1}^{U_{n}}x_{nmu}(t)p_{nmu}(t)h_{nmu}(t)\leq I_{m}^{S}(t)
,\forall m \\
& C5: \sum\limits_{n=1}^{N}\sum\limits_{u=1}^{U_{n}}x_{nmu}(t)\leq 1 ,\forall n,m \\
& C6: x_{nmu}(t)\in \left\{ 0,1 \right\} ,\forall n,m,u \\
& C7: 0 \le {F_n}\left( t \right) \le {S_n}\left( t \right)\\
& C8: 0\leq J_n(t)\leq \min\left \{ S_n^{max}-S_n(t),E_n(t) \right \}\\
& C9: S_{n}(t+1)=\left[ S_{n}(t)-F_{n}(t)\right]^+ + J_{n}(t)\\
& C10: \text{Queues }Q_{n,u}\left(t\right) \text{are strongly stable}, \forall n,u
\end{split}
\end{equation}
where variables $\mathbf{D}=\left\{D_{nu}(t)\right\}$, $\mathbf{X}%
=\left\{x_{nmu}(t)\right\}$, $\mathbf{P}=\left\{p_{nmu}(t)\right\}$, $%
\mathbf{G}=\left\{G_{n}(t)\right\}$ are admitted data rate, subchannel
assignment, power control and grid energy consumption variables of the
network respectively. $C1$ ensures that SBSs will benefit from sharing spectrum resources
and $C_{n}^{min}$ is the minimal profit of SBS $n$ in the non-cooperative
way, which is set to be a time-invariant non-negative constant. $C2$-$C9$ are the admitted data amount, subchannel, power and energy constraints. $C10$ is imposed to ensure the network stability.

As for the objective function in optimization problem OP$_1$, since the $log$ is a continuous strictly increasing function, $\max\sum_{n=1}^{N}\log(\overline{C}_{n}-{C_{n}^{min}})$ is equivalent to $\max\prod_{n=1}^{N}(%
\overline{C}_{n}-{C_{n}^{min}})$, which is the Nash product. Based on the Nash Bargaining Solution (NBS), by maximizing the Nash product, or equivalently the $log$ sum in optimization problem OP$_1$, a unique and fair Pareto optimality solution can be provided \cite{29}.

According to the energy queue model, we can find the current energy
management policy has a coupling relationship with future energy state. The
coupling nature, which comes with stochastic renewable energy sources, makes
$\mathrm{OP}_1$ more complicated. By taking iterated expectation and using
telescoping sums in (14) over $t=0,1,...,T-1$, we can get
\begin{equation}
\mathbb{E}\left\{ S(T-1) \right\} -S(0) = \sum_{t=0}^{T-1}\mathbb{E}\left \{
-F_{n}(t)+J_n(t) \right \}\label{eq-23}
\end{equation}

As the energy level of the battery devices $S_n(t)$ is bounded by $S_n^{max}$
, we have $\overline{F}_{n}=\overline{J}_{n}$ after dividing both sides
of \eqref{eq-23} by $T$ and taking limitation of  $T\rightarrow \infty $.

Moreover, it is noted that, the optimization problem $\text{OP}_1$ aims to maximize an objective function of a time-averaged variable $\overline{C}_n-{C_n^{min}}$, which is difficult to be tackled. To solve $\text{OP}_1$, we introduce an auxiliary variable $\mu _{n}$ to reformulate problem $\mathrm{OP}_1$ as
\begin{equation}
\begin{split}
\mathrm{OP}_2: &\quad\max\limits_{\mathbf{D,X,P,G},\boldsymbol{\alpha,\beta}}\sum\limits_{n=1}^{N}\log(%
\overline{\mu _{n}})\\
\text{s.t.}\quad &C1^{\prime }:\overline{\mu} _{n}\leq \overline{C}_{n}-{C_{n}^{min}} \\
&C2^{\prime }:\overline{C}_{n}\geq C_{n}^{min},\forall n \\
&C3^{\prime }:C2-C6 \\
&C4^{\prime }:\overline{F}_{n}=\overline{J}_{n} \\
&C5^{\prime }:\text{Queues }Q_{n,u}\left(t\right) \text{are strongly stable}, \forall n,u
\end{split}
\end{equation}
where $\mu _{n}$ can be understood as the lower bound of $\overline{C}_{n}-%
{C_{n}^{min}}$ as shown in $C1^{\prime }$. Based on the Jensen inequality, %
$\overline{\log(\mu _{n})}$ is the lower bound of $\log(\overline\mu _{n})$ with the non-decreasing concave logarithmic utility function. Thus, it is feasible to change the objective function $\log(\overline\mu _{n})$ to $\overline{\log(\mu _{n})}$, i.e., the optimization problem $\text{OP}_2$ is an optimization problem with a time-averaged objection function, which is beneficial for the following problem solving.

To satisfy the time-average constraints $C1^{\prime }$ and $C2^{\prime }$, we introduce the
concept of virtual queue technology\cite{27}, and use virtual queues of $Y(t)$ and $Z(t)$.
Specifically, the virtual queue $Y(t)$ associated with
constraint $C1^{\prime }$ updates as follows:
\begin{equation}
Y_{n}(t+1)=\left[Y_{n}(t)-y_{n}^{out}(t)\right]^++y_{n}^{in}(t)
\end{equation}
where
\begin{equation}
y_{n}^{in}(t)=\mu _{n}(t)+C_{n}^{min}+\phi G_{n}(t)
\end{equation}
\begin{equation}
y_{n}^{out}(t)=\sum_{u=1}^{U_{n}}{D}_{nu}(t)+O_n(t)
\end{equation}

The virtual queue $Z(t)$ associated with constraint $C2^{\prime }$ updates as
follows:
\begin{equation}
Z_{n}(t+1)= \left[Z_{n}(t)-z_{n}^{out}(t)\right]^++z_{n}^{in}(t)
\end{equation}
where
\begin{equation}
z_{n}^{in}(t)=C_{n}^{min}+\phi G_{n}(t)
\end{equation}
\begin{equation}
z_{n}^{out}(t)=\sum_{u=1}^{U_{n}}{D}_{nu}(t)+O_n(t)
\end{equation}

If the virtual queues of $Y_{n}(t)$ and $Z_{n}(t)$ are strongly stable, then the
constraints $C1^{\prime }$ and $C2^{\prime }$ are satisfied.

\subsection{Lyapunov Optimization}
According to the queues, $\mathbf{Q}=\left \{Q_{nu}(t) \right \}$, $\mathbf{S}=\left
\{S_{n}(t) \right \}$, $\mathbf{Y}=\left \{Y_{n}(t) \right
\}$, $\mathbf{Z}=\left \{Z_{n}(t) \right \}$, and $\Theta (t)=[\mathbf{Q},%
\mathbf{S},\mathbf{Y},\mathbf{Z}]$. We define the Lyapunov function as
\begin{equation}  \label{equ:lyap}
\begin{split}
L(\Theta (t))=\frac{1}{2}\left[\sum\limits_{n=1}^{N}\sum%
\limits_{u=1}^{U_{n}}Q_{nu}(t)^{2}+\sum\limits_{n=1}^{N}\left(
S_{n}(t)-\rho_n \right)^2+\sum\limits_{n=1}^{N}Y_{n}^2(t)+\sum\limits_{n=1}^{N}Z_{n}^2(t)\right]
\end{split}%
\end{equation}
where $\rho_n$ is a perturbation factor, which ensures there is enough
energy in the energy queue. Without loss of generality, we assume that all queues are empty when $t = 0$ such that $L(\Theta (t))=0$.

The Lyapunov function $L(\Theta (t))$ is a scalar measure of network
congestion. Intuitively, if $L(\Theta (t))$ is small then all queues are
small; and if $L(\Theta (t))$ is large then at least one queue is large.
Thus, by minimizing the drift in the Lyapunov function (i.e., by minimizing a
difference in the Lyapunov function from one time slot to the next), queues $%
Q_{nu}(t)$, $S_{n}(t)$, $Y_{n}(t)$, $Z_{n}(t)$ can be stabilized. By using
expression (\ref{equ:lyap}), the drift in the Lyapunov function (i.e., the
expected change in the Lyapunov function from one time slot to the next) can be
written as
\begin{equation}
\bigtriangleup (\Theta (t))=\mathbb{E}\left \{ L(\Theta (t+1))-L(\Theta
(t))|\Theta (t) \right \}
\end{equation}

We now use the drift-plus-penalty minimization method to solve optimization problem OP$_2$.
In this method, a control policy that solves OP$_2$ is obtained by
minimizing the upper bound on the following drift-plus-penalty expression. We
can obtain the following drift-plus-penalty term:
\begin{equation}
\Delta _V(t)=\bigtriangleup (\Theta (t))\underset{\mathrm{penalty\ term}}{%
\underbrace{-V\sum_{n=1}^{N}\mathbb{E}\left \{ f(\mu _{n}(t))|\Theta (t)
\right \}}}
\end{equation}
%\begin{equation}
%\Delta _V(t)=\bigtriangleup (\Theta (t))-V\sum_{n=1}^{N}\mathbb{E}\left \{ log(\mu _{n}(t))|\Theta (t) \right \}
%\end{equation}
where $V$ is a non-negative tunable parameter which balances the maximization
of network utility and the minimization of queue length to a state level.
According to \cite{30}, when $V$ is sufficiently large, the optimization
algorithm approaches the optimal capacity.

\textbf{Lemma 1.} The drift-plus-penalty term is upper bounded as
\begin{align}
& \Delta _V(t)\leq H+\sum\limits_{n=1}^{N}\sum\limits_{u=1}^{U_{n}}Q_{nu}(t)%
\mathbb{E}\left \{ D_{nu}(t)-R_{nu}(t)|\Theta (t) \right \}  \notag \\
&+ \sum\limits_{n=1}^{N}W_{n}(t)\mathbb{E}\Bigg\{C_{n}^{min} +\phi
G_{n}(t)-\sum_{u=1}^{U_{n}}{D}_{nu}(t)-O_n(t)|\Theta (t) \Bigg\}  \notag \\
&+\sum\limits_{n=1}^{N}\Bigg\{\left( S_{n}(t)-\rho_n \right)\mathbb{E}\left
\{ J_{n}(t)-P_{n}(t)+G_n(t)|\Theta (t) \right \}  \notag \\
&+Y_{n}(t)\mathbb{E}\left \{ \mu _{n}(t)|\Theta (t)\right \}\Bigg\} %
-V\sum_{n=1}^{N}\mathbb{E}\left \{ f(\mu _n(t))|\Theta (t) \right \}
\label{equ:lya}
\end{align}
where $W_n(t)=Y_n(t)+Z_n(t)$ denotes the value of virtual queues. $H$ is a
positive constant that satisfies the following inequality constraint for all
time slots:
\begin{align}
H\geq &\frac{1}{2} \Bigg[ \sum\limits_{n=1}^{N}\sum%
\limits_{u=1}^{U_{n}}A_{max}^2+\sum\limits_{n=1}^{N}J_n^{max2}+\sum\limits_{n=1}^{N}S_n^{max2}+\sum\limits_{n=1}^{N}\sum%
\limits_{u=1}^{U_{n}}\mathbb{E}\left \{R_{nu}(t)^2|\Theta (t) \right \}
\notag \\
&+\sum\limits_{n=1}^{N}\mathbb{E}\left \{
y_{n}^{in}(t)^2+y_{n}^{out}(t)^2|\Theta (t) \right \}+\sum\limits_{n=1}^{N}\mathbb{E}\left \{
z_{n}^{in}(t)^2+z_{n}^{out}(t)^2|\Theta (t) \right \} \Bigg]
\end{align}

\textit{proof:} See Appendix A.

By \textit{Lemma 1}, we have transformed the optimization problem OP$_2$ into
minimizing the right-side term of (\ref{equ:lya}) at each time slot $t$.
According to \cite{26}, the control policy should be adjusted to minimize
the upper bound. Thus, we will decompose the optimization problem and
present an online dynamic control algorithm for the green resource allocation and energy management.

\section{Online Control Algorithm for SBSs}

In this section, we focus on the online algorithm design based on
the previous subsection. In the multi-person bargaining problem, SBSs will
compete with the total spectrum band. We propose an online dynamic control
algorithm by drift-plus-penalty method in Algorithm 1. At each time slot, the Algorithm 1
is partitioned to four steps:

1) Admitted data rate control: each SBS decides the admission rate of each
SUE according to the data queue length and the virtual queue length;

2) Auxiliary variable decision: each SBS decides the auxiliary variable $%
\mu_{n}(t)$;

3) Spectrum pricing and resource allocation: each SBS decides either lease
or rent the spectrum band; and each SBS designs subchannel assignment and
power allocation scheme;

4) Battery energy management: each SBS decides the amount of discharge
energy and charge energy to reduce the grid energy consumption according to
the perturbation variable.

\begin{algorithm}
	\caption{Online Dynamic Control Algorithm for Green Resource Allocation and Energy Management}
	\begin{algorithmic} \label{algm1}
		\STATE \textbf{Initial Stage:}
		\STATE At the beginning of all time slot, initialize all the queues $Q_{nu}(0),S_{n}(0),Y_{n}(0),Z_{n}(0)=0$

		\STATE \textbf{Step 1--Admission rate control:}
		\STATE For each SBS $n$, calculate the flow rate by solving the problem in (\ref{equ:arc}).

		\STATE \textbf{Step 2--Auxiliary variable decision:}
		\STATE For each SBS $n$, calculate the variable $\mu_{n}(t)$ by solving the problem in (\ref{equ:avd}).

		\STATE \textbf{Step 3--Adaptive resource allocation:}
		\STATE For each SBS $n$, compute spectrum sharing variables $\alpha_n(t)$, $\beta_n(t)$ by solving the problem in (40) and resource allocation variables $p_{nmu}(t)$, $x_{nmu}(t)$ by solving the problem in (43).
		\STATE \textbf{Step 4--Battery energy management:}
		\STATE For each SBS $n$, calculate the discharge energy, charge energy and grid energy by solving the problem in (\ref{equ:bem}).

		\STATE \textbf{End Stage:}
		\STATE Update the queue state and go to \textbf{Step 1}.
	\end{algorithmic}
\end{algorithm}

\subsection{Admission Rate Control}

By observing the second and third terms on the right-hand side of (\ref%
{equ:lya}), we can get the rate control problem regardless of the other variables:
\begin{align}  \label{equ:arc}
\max W_{n}(t)D_{nu}(t)-Q_{nu}(t)D_{nu}(t) \\
\text{s.t.} \quad 0 \leq D_{nu}(t)\leq A_{nu}(t), \forall n,u,t  \notag
\end{align}

We can observe the monotonicity (\ref{equ:arc})
and then the solution of the above problem can be solved easily,
\begin{align}
D_{nu}(t)=\left\{%
\begin{matrix}
A_{nu}(t), & W_n(t)\geq Q_{nu}(t) \\
0, & W_n(t)<Q_{nu}(t)%
\end{matrix}%
\right.
\end{align}

\subsection{Auxiliary Variable Decision}

Each SBS decides the auxiliary variable $\mu_{n}(t)$ by
\begin{align}  \label{equ:avd}
\max V\log(\mu_{n}(t))-Y_{n}(t)\mu_{n}(t) \\
\mathrm{s.t.} \quad 0 \leq \mu_{n}(t)\leq \mu_n^{max}, \forall n,t  \notag
\end{align}

$\log(x)$ is the non-decreasing concave utility function, and then we get the
optimal solution. $\mu_n^{max}$ is the maximal profit by sharing spectrum
resources.
\begin{align}  \label{equ:avdd}
\mu_{n}(t)=\left\{%
\begin{matrix}
\mu_n^{max}, & Y_n(t)\leq V\log^{\prime }(\mu_n^{max}) \\
\mu_n^*, & V\log^{\prime }(\mu_n^{max})<Y_n(t)< V\log^{\prime }(t) \\
0, & Y_n(t)\geq V\log^{\prime }(t)%
\end{matrix}%
\right.
\end{align}
where $\mu_n^*$ should satisfy $V\log^{\prime }(\mu_n^*)=Y_n(t)$.

\subsection{Dynamic Resource Allocation}

We can formulate the pricing problem as {\small \
\begin{align}
\max \sum_{n=1}^{N} W_{n}(t)\Big( &\alpha_{n}(t)\left ( B_{n}^0-B_n(t)
\right )^+-\beta_{n}(t)\left ( B_n(t)-B_{n}^0 \right )^+ \Big)  \notag \\
&\text{s.t.} \quad 0 \leq \alpha_{n}(t),\beta_{n}(t)\leq q_n^{max}
\end{align}
}

We can find that the subchannel price $\alpha_{n}(t)$ and $\beta_{n}(t)$ are
related to the state of virtual queue $W_n(t)$. The longer the queue length
of $W_n$, the more utility gain the SBS $n$ will obtain by renting its free
spectrum resources and vice versa. Moreover, it is easily seen that
the pricing problem is tightly coupled with the power and subchannel allocation problem.
In the following, to facilitate problem solving, we firstly analyze the resource allocation in two-SBS case,
and then extend the results to the multi-SBS case.

%{Due to the autonomy and distributed characters of SBSs, they can make rational and independent decisions. Therefore, each SBS will pursue maximum profit, which causes a conflict of interest. However, SBSs also cooperate with each other to maximize the network profits. First, we solve the resource allocation problem for two SBSs, and then extend it to a multi-SBSs problem.}

1) Two-SBS case ($N=2$): According to the spectrum pricing problem, the SBS
with the longer queue length of $W_n$, which is supposed the SBS 1, will
rent spectrum resources to another one (SBS 2). The price is chosen as
\begin{equation}
\left [\alpha,\beta \right ]=\left\{%
\begin{matrix}
\alpha_{1}(t)=q_1^{max},\beta_{1}(t)=0 \\
\alpha_{2}(t)=0,\beta_{2}(t)=q_2^{max}%
\end{matrix}%
\right.
\end{equation}

By observing the remaining term on the right-hand side of (\ref{equ:lya}),
we can formulate the resource allocation problem with the given spectrum
sharing scheme as follows:
\small{\begin{equation}
\begin{split}
\underset{\mathbf{p,x}}{\max} \sum\limits_{n=1}^{N}\sum%
\limits_{u=1}^{U_{n}}&Q_{nu}(t)R_{nu}(t)+\sum\limits_{n=1}^{N}W_{n}(t)O_n(t)+\sum\limits_{n=1}^{N}\left(S_{n}(t)-\rho_{n}\right)P_{n}(t)\\
\text{s.t.}\ &C3: \sum\limits_{m=1}^{M}\sum\limits_{u=1}^{U_{n}}x_{nmu}(t)p_{nmu}(t)\leq
p_n^{max}, \forall n \\
&
C4:\sum\limits_{n=1}^{N}\sum%
\limits_{u=1}^{U_{n}}x_{nmu}(t)p_{nmu}(t)h_{nmu}(t)\leq I_{m}^{S}(t), \forall m \\
& C5: \sum\limits_{n=1}^{N}\sum\limits_{u=1}^{U_{n}}x_{nmu}(t)\leq 1,  \forall m\\
& C6: x_{nmu}(t)\in \left [ 0,1 \right ], \forall n,m,u\\
& C11: p_{nmu}(t)\geq 0, \forall n,m,u\\
\end{split}\label{eqo:ras}
\end{equation}}

We relax the binary variable $x_{nmu}(t)$ to a continuous variable $\hat{x}%
_{nmu}(t)\in \left[0,1\right]$. For notational brevity, denote the power
allocated to SUE $u$ on subchannel $m$ as $s_{nmu}=\hat{x}_{nmu}p_{nmu}$.
The optimization problem can be rewritten as:
\begin{equation}
\begin{split}
\underset{\mathbf{p,x}}{\max}&\sum\limits_{n=1}^{N}\sum\limits_{m=1}^{M}%
\sum\limits_{u=1}^{U_{n}}\Bigg( Q_{nu}(t)\varpi_m\hat{x}_{nmu}(t)\times \log_2 \left( {1 + \frac{{{s_{nmu}}\left( t \right){h_{nmu}}\left( t \right)}}{{\hat{x}_{nmu}(t)\left( {{I_{0mu}}\left( t \right) + {\sigma ^2}} \right)}}} \right) +\eta_{n}(t)s_{nmu}(t)+\theta _{nm}(t)\hat{x}_{nmu}(t)\Bigg)\\
\text{s.t.} \ &\bar{C3}: \sum\limits_{m=1}^{M}\sum\limits_{u=1}^{U_{n}}s_{nmu}(t)\leq
p_n^{max},\forall n \\
& \bar{C4}:\sum\limits_{n=1}^{N}\sum\limits_{u=1}^{U_{n}}s_{nmu}(t)h_{nmu}(t)\leq
I_{m}^{S}(t),\forall m \\
& \bar{C5}: \sum\limits_{n=1}^{N}\sum\limits_{u=1}^{U_{n}}\hat{x}_{nmu}(t)\leq 1
,\forall m \\
& \bar{C6}: \hat{x}_{nmu}(t)\in \left [ 0,1 \right ]\forall n,m,u\\
& \bar{C11}:s_{nmu}(t)\geq 0,\forall n,m,u \\
\end{split}\label{ali:pa}
\end{equation}
where $\theta _{nm}(t)$ is related to $\alpha_{n}(t)$, $\beta_{n}(t)$, $%
\varpi_m$ and $W_{n}(t)$, ${\theta _{nm}}\left( t \right) =  - q_n^{\max }{\varpi _m}{W_n}\left( t \right)$ and $\eta _n(t)=S_n(t)-\rho_n-\phi W_n(t)$. $%
I_{0mu}=p_{mk}^M(t)h_{nmu}^M(t)$. The subchannel and power allocation
strategy in (\ref{ali:pa}) can be solved by using the Lagrangian dual
decomposition method. By ignoring the time variables, the partial Lagrangian
function is given by
\begin{align}  \label{ali:lag}
L(x&,s
,\lambda_1,\lambda_2,\lambda_3)=\sum\limits_{n=1}^{N}\sum%
\limits_{m=1}^{M}L_{nm}(x,s ,\lambda_1,\lambda_2,\lambda_3){\rm{ + }}\sum\limits_{n = 1}^N {{\lambda _{1,n}}p_n^{\max } + } \sum\limits_{m = 1}^M {{\lambda _{2,m}}I_m^S + } \sum\limits_{m = 1}^M {{\lambda _{3,m}}}
\end{align}
with
\begin{align}
L_{nm}(x,s ,\lambda_1,\lambda_2,\lambda_3)=&\sum\limits_{u=1}^{U_{n}}\Bigg[ %
Q_{nu}(t)\varpi_m\hat{x}_{nmu}(t)\times \log_2 \left( {1 + \frac{{{s_{nmu}}\left( t \right){h_{nmu}}\left( t \right)}}{{\hat{x}_{nmu}(t)\left( {{I_{0mu}}\left( t \right) + {\sigma ^2}} \right)}}} \right)  \notag \\
&+\eta_{n}s_{nmu}(t)+\theta _{nm}(t)\hat{x}_{nmu}(t)\Bigg] -\lambda_{1,n}\sum\limits_{u=1}^{U_{n}}s_{nmu}(t) \notag \\&-\lambda_{2,m}\sum\limits_{u=1}^{U_{n}}s_{nmu}(t)h_{nmu}(t)
-\lambda_{3,m}\sum%
\limits_{u=1}^{U_{n}}\hat{x}_{nmu}(t)\notag
\end{align}
where $\lambda_1$, $\lambda_2$ and $\lambda_3$ are the Lagrange multipliers for constraints
$\bar{C3}$, $\bar{C4}$, $\bar{C5}$ in (\ref{ali:pa}) respectively. The boundary constraints $\bar{C5}$
and $\bar{C11}$ will be absorbed in the Karush-Kuhn-Tucker (KKT) conditions. Thus,
the Lagrangian dual function is defined as:
\begin{equation}
g(\lambda)=\max\limits_{x,s}L(x,s,\lambda_1,\lambda_2,\lambda_3)
\end{equation}
The dual problem can be expressed as:
\begin{equation}
\begin{split}
\min\limits &\quad g(\lambda_1,\lambda_2,\lambda_3)\\
\text{s.t.} &\quad\lambda_1,\lambda_2,\lambda_3\geq 0
\end{split}
\end{equation}

According to the KKT conditions, the optimal solutions of the problem
should satisfy the following conditions:
\begin{align*}
\frac{\partial L_{nm}(t)}{\partial s_{nmu}(t)}&=\frac{1}{\ln2}%
\frac{Q_{nu}(t)\varpi_m\hat{x}_{nmu}(t)}{s_{nmu}(t)+\frac{\hat{x}_{nmu}(t)\left( {{I_{0mu}}\left( t \right) + {\sigma ^2}} \right)}{h_{nmu}(t)}}+\eta_n-\lambda _{1,n} -\lambda_{2,m}h_{nmu}(t)=0\\
&\hat{x}_{nmu}(t)\in \left [ 0,1 \right ], s_{nmu}(t) \geq 0
\end{align*}
%\begin{equation*}
%\varrho _{nmu}\geq 0
%\end{equation*}
%where $\varrho _{nmu}$ is the Lagrange multiplier for the constraint $p_{nmu}\geq 0$.

Thus, we can get the optimal power allocation:
\begin{equation}
s_{nmu}^*(t)=\left [ \frac{1}{\ln2}\frac{\varpi_m Q_{nu}(t)}{\omega _{nmu}-\eta_{n}}%
-\frac{I_{0mu}(t)+\sigma^2}{h_{nmu}(t)} \right ]^+\hat{x}_{nmu}(t)
\end{equation}
where $\omega _{nmu}=\lambda_{1,n}+\lambda_{2,m}h_{nmu}(t)$

Then, we will make use of the results of power allocation for subcarrier assignment. We decompose (45) into $U_n$ independent subproblems. Each subproblem is formulated as following:
\begin{equation}
{L_{nm}}(\mathbf{P}) = \sum\limits_{u = 1}^{{U_n}} {{L_{nmu}}(\mathbf{P})}
\end{equation}
where
\begin{align}
L_{nmu}(\mathbf{P})=&Q_{nu}(t)\varpi_m\hat{x}_{nmu}(t)\times\log_2\left(1+\frac{s^*_{nmu}(t)h_{nmu}(t)}{%
\hat{x}_{nmu}(t)(I_{0mu}(t)+\sigma ^{2})} \right)+\theta _{nm}\hat{x}_{nmu}(t)+\eta_{n}s^*_{nmu}(t)\notag\\
&-\lambda_{1,n}s^*_{nmu}(t)-\lambda_{2,m}s^*_{nmu}(t)h_{nmu}(t) -\lambda_{3,m}\hat{x}_{nmu}(t)
\end{align}
Substituting (48) into (50), the objective of subcarrier assignment is to maximize $L_{nm}(\mathbf{P})$ for all SUEs associated with SBS $n$.
For any subcarrier $m$, it will be assigned to the SUE who has
the biggest $L_{nmu}(\mathbf{P})$. Let $m^*_u$ be the result of subcarrier  $m$'s assignment, which is given by:
\begin{equation}
m_u^* = \mathop {\arg \max }\limits_u {L_{mnu}}\text{ and }{L_{m_u^*nu}} > 0
\end{equation}
\begin{equation}
x_{nmu}^*(t)=\left\{%
\begin{matrix}
1 & \text{if} \quad m=m^*_u \\
0 & \text{otherwise}%
\end{matrix}%
\right.
\end{equation}

We use the subgradient method to update the Lagrange multipliers:
\begin{equation}
\lambda_{1,n}^{(i+1)}=\Bigg[\lambda_{1,n}^{(i)}-d_{1}^{(i)}(p_n^{max}-\sum%
\limits_{m=1}^{M}\sum\limits_{u=1}^{U_{n}}s_{nmu}(t))\Bigg]^{+}
\end{equation}
\begin{equation}
\lambda_{2,m}^{(i+1)}=\Bigg[\lambda_{2,m}^{(i)}-d_{2}^{(i)}\left (
I_{m}^{S}-\sum\limits_{n=1}^{N}\sum\limits_{u=1}^{U_{n}}s_{nmu}(t)h_{nmu}(t)
\right )\Bigg]^{+}
\end{equation}
\begin{equation}
\lambda_{3,m}^{(i+1)}=\Bigg[\lambda_{3,m}^{(i)}-d_{3}^{(i)}(1-\sum%
\limits_{n=1}^{N}\sum\limits_{u=1}^{U_{0}}x_{nmu}(t))\Bigg]^{+}
\end{equation}
where $d_1^{(i)},d_2^{(i)},d_3^{(i)}$ are the step sizes of iteration $i$.

Finally, we obtain the proposed spectrum pricing and resource allocation
algorithm.

It is noted that, according to \cite{33}, the convergence of Eqs. (53)-(55) can be guaranteed by adopting diminishing step sizes $d_k^{(i)}, k=1,2,3$. Moreover, since only Step 3 in Algorithm 1 involves iterations. Therefore, the convergence of the proposed algorithm is determined by Eqs. (53)-(55), and by adopting diminishing step sizes $d_k^{(i)}, k=1,2,3$, the convergence of the proposed Algorithm 1 can be guaranteed. On the other hand, in practice, the subproblem (47) is solved by each SBS locally for $MU_n$ times during one iteration $i$ and the computational complexity at each SBS is $O(MU_n)$.

2) Multi-SBS case ($N>2$): For the spectrum sharing of multiple SBSs, we decompose the original
problem into several two-SBS problems. We first group SBSs into multiple pairs and
then use dynamic resource allocation in two-SBS case for each pair. Notice that if the
number of SBSs is odd, we can set a virtual SBS with average indexes
determined by historical data. From this, we can establish a SBS-pair
problem:
\begin{equation}
\begin{split}
\max &\quad\sum\limits_{i=1}^N\sum\limits_{j=1}^N a_{ij}\widetilde{C}_{ij}\\
\text{s.t.} &\sum_{i=1}^{N}a_{ij}=1,\sum_{j=1}^{N}a_{ij}=1,a_{ij}\in \left \{
0,1 \right \}
\end{split}
\end{equation}
where $a_{ij}$ is the pair parameter. If SBS $i$ and SBS $j$ form a spectrum
sharing pair, $a_{ij}=1$. Otherwise, $a_{ij}=0$. $\widetilde{C}_{ij}$ is the
relative benefit for the SBS $i$ sharing spectrum with SBS $j$, which is
compared with the payoff before sharing. The pairing issues can be solved by
Hungarian method \cite{34}. Moreover, the complexity of the Hungarian method is $O(N^4)$.
Thus, the overall complexity for each iteration of the proposed online dynamic control algorithm
in multi-SBS case is $O(MU_n+N^4)$.

\subsection{Battery Energy Management}

By observing the third term on the right-hand side of (\ref{equ:lya}), we
can get the harvested energy management problem for each SBS $n$:
\begin{align}  \label{equ:bem}
\min\,\,&(S_{n}(t)-\rho_{n}+\phi W_n(t))G_{n}(t)+(S_n(t)-\rho_n)J_n(t)  \notag
\\
\text{s.t.}\,\,&P_n(t)=F_n(t)+G_n(t), \forall n,t  \notag \\
&0 \le {F_n}\left( t \right) \le {S_n}\left( t \right) \\
&0\leq J_n(t)\leq \min\left \{ S_n^{max}-S_n(t),E_n(t) \right \}\notag
\end{align}

The solution of (\ref{equ:bem}) consists of three situations:

a) $S_n(t) > \rho_n$

The storage device does't harvest renewable energy and provide main energy
for transmission. Thus, we have the optimal energy management scheme:
\begin{align}
\left\{%
\begin{matrix}
J_n^*(t) & = & 0 \\
F_n^*(t) & = & \min\left \{ P_n^*(t),S_n(t) \right \} \\
G_n^*(t) & = & 0%
\end{matrix}%
\right.
\end{align}

b) $\rho_n-\phi W_n(t) < S_n(t) \leq \rho_n$

The SBS harvests energy to feed the battery and the battery provides main
energy for transmission. The energy harvest and power supply scheme is
\begin{align}
\left\{%
\begin{matrix}
J_n^*(t) & = & \min\{S_n^{max}-S_n(t),E_n(t)\} \\
F_n^*(t) & = & \min\left \{ P_n^*(t),S_n(t) \right \} \\
G_n^*(t) & = & 0%
\end{matrix}%
\right.
\end{align}

c) $S_n(t) \leq \rho_n-\phi W_n(t)$

The battery level is inadequate for normal transmission then the SBS will be
supplied by the grid. The energy scheduling scheme is
\begin{align}
\left\{%
\begin{matrix}
J_n^*(t) & = & \min\{S_n^{max}-S_n(t),E_n(t)\} \\
F_n^*(t) & = & \min\left \{ P_n^*(t),S_n(t) \right \} \\
G_n^*(t) & = & \max\left \{ 0,P_n^*(t)- F_n^*(t)\right \}%
\end{matrix}%
\right.
\end{align}

\subsection{Discussions on Algorithm Implementation}
It is noted that, in the spectrum pricing and resource allocation for multi-SBS case ($N>2$), the Hungarian method-based pairing may make the proposed algorithm be inefficient for a large-scale or dense LTE/5G network. Then, the zoning or clustering scheme in LTE can be adopted to decrease the complexity of the proposed scheme \cite{20,35,36,37}. Particularly, within a zone, several SBSs bargain on spectrum sharing and then perform the spectrum pricing and resource allocation.

\begin{figure}[htbp]
\centering
\includegraphics[width=0.6%
\textwidth]{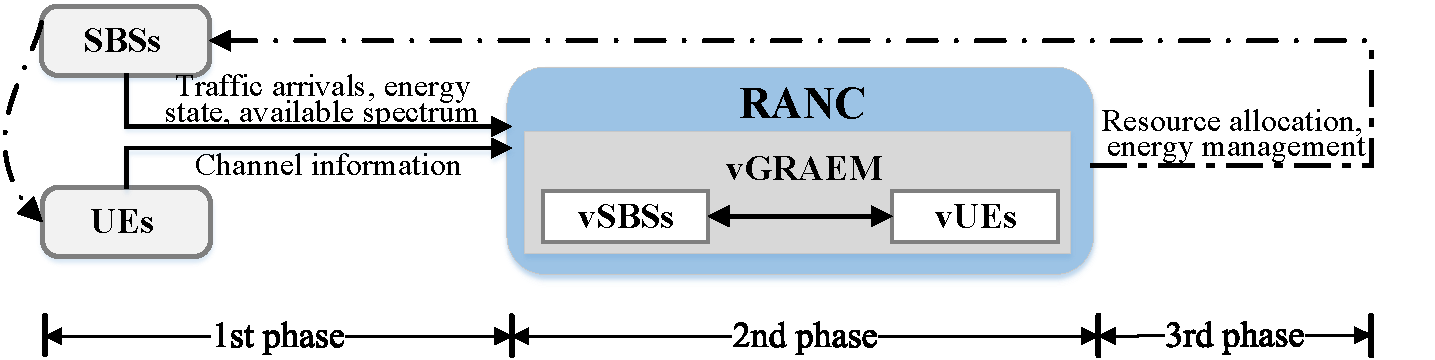}\newline
\caption{Implementation of vGRAEM}\label{fig:vgraem}
\end{figure}

On the other hand, based on the newly emerged software-defined networking (SDN) architecture, which can accelerate the innovations for both hardware forwarding infrastructure and software networking algorithms through control and data separation, enable efficient and adaptive sharing of network resources, and achieve maximum spectrum efficiency and enhance energy efficiency \cite{38}, we can designe a virtual green resource allocation and energy management (vGRAEM) scheme, which will reduce the communication overhead over air interface and avoid the information leaking to UEs. Mainly, by leveraging cloud computing and network virtualization, virtual UEs (vUEs) and virtual SBSs (vSBSs) can be generated in the radio access networks controller (RANC) to emulate a resource allocation and energy management solution. The implementation of vGRAEM is shown in Fig. \ref{fig:vgraem}. Generally, it consists of three phases. The 1st phase is the initial network measurements; in the 2nd phase, the RANC first generates vUEs and vSBSs and then simulates the dynamic control algorithm for resource allocation and energy management based on the information collected in the 1st phase; in the 3rd phase, the RANC informs individual SBSs about resource allocation and energy management decisions.

\section{Performance Analysis}

In this section, we analyse the properties of the queues and proposed online dynamic control algorithm.

\textit{Theorem 1:} By setting the perturbation $\rho_n$ as
\begin{align}
\rho_{n}=S_n^{max}-E_n^{max}
\end{align}
the energy queue is bounded by $0\leq S_n(t)\leq S_n^{max}$.

\textit{proof:} The proof can be found in Appendix B.

\textit{Theorem 2:} a) We define ${f\left( {\mu _n^*\left( t \right)} \right)}$ as our optimal decision and suppose there exist finite and positive constants $\varepsilon$ and $V$. The proposed online dynamic control algorithm ensures that the data queue and virtual queues have an upper bound.
\begin{align}
\underset{T\rightarrow \infty }{\mathrm{lim}}&\mathrm{sup}\frac{1}{T}\sum_{t=0}^{T-1}\mathbb{E}\left [ {\sum\limits_{n = 1}^N {\sum\limits_{u = 1}^{{U_n}} {{Q_{nu}}\left( t \right)} }  + \sum\limits_{n = 1}^N {{Y_n}\left( t \right) + \sum\limits_{n = 1}^N {{Z_n}\left( t \right)} } } \right ] \nonumber\\
&\leq \frac{{{H} + V\left( {\overline f  - {f^*}} \right)}}{{\varepsilon}}
\end{align}

b) The proposed algorithm can achieve the near optimal capacity

\begin{equation}
\overline f  \ge {f^*} - \frac{{{H}}}{V}
\end{equation}
where \small{$\overline f\!\!=\!\!\lim \mathop {\sup }\limits_{T \to \infty } \frac{1}{T}\sum\limits_{t = 0}^{T - 1} {V\mathbb{E}\left\{ {\sum\limits_{n = 1}^{N} {f\left( {\mu \left( t \right)} \right)} } \right\}}$}, ${f^*}\!\!=\!\!{\sum\limits_{n = 1}^N {f\left( {\mu _n^*\left( t \right)} \right)} } $

\textit{proof:} The proof can be found in Appendix C.

\section{Simulation Results}

\begin{figure}[h]
\centering
\includegraphics[width=7cm]{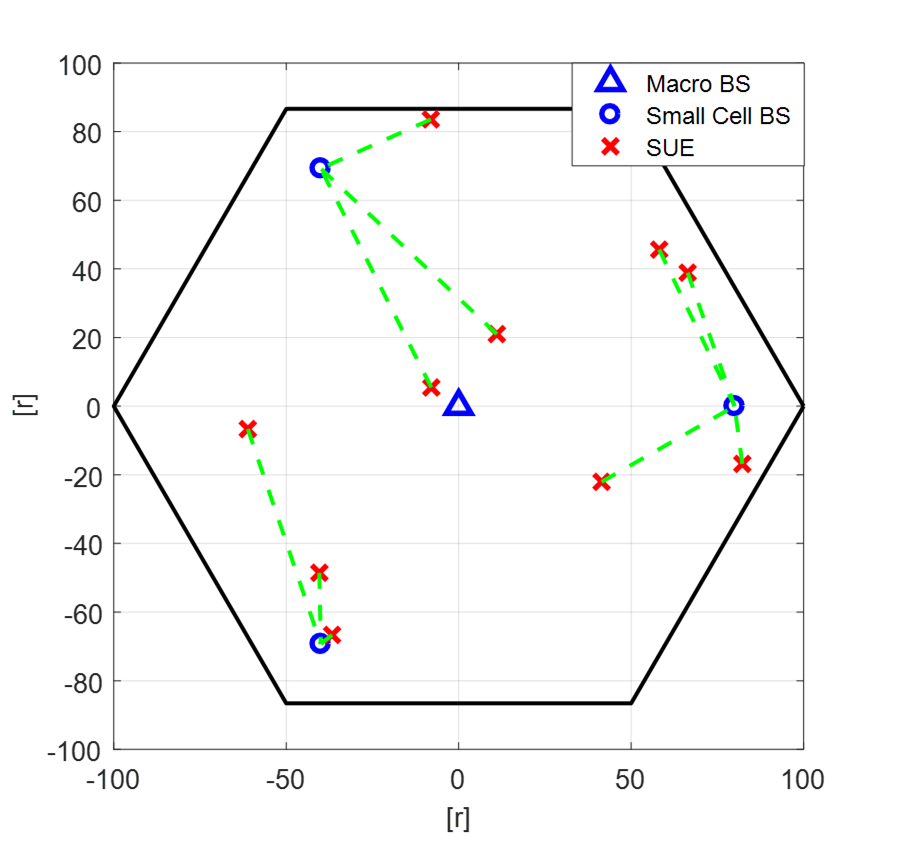}\newline
\caption{A macrocell with three coexisting SBSs}\label{fgi:simumod}
\end{figure}

We simulate a HetNet where a MBS is underlaid with three uniformly
distributed SBSs. The SUEs are randomly distributed in the coverage of their
serving SBSs as shown in Fig. \ref{fgi:simumod}. The total bandwidth is 30 MHz and we suppose
the SBSs divide the spectrum band equally in the initial state. $p_n^{max}$%
=0.1 W, $I_{th}^S=2*10^{-10}$, $\sigma^2=BN_0$ where $N_0=-174$ dBm/Hz is
the AWGN power spectral density. The
capacity of the SBS storage device is 500 Wh. We simulate the channel as path
loss, Rayleigh fading and shadowing effect with mean zero and deviation
10dB. The MBS transmits at its peak power $P_m=40$dBm and transmits on each
subchannel at the same power. The data arrival is subject to poisson
distribution with the mean value 4 packets/slot. The mean packet size is
5000 bits/packet. The static power consumption is 3.2W and the power
conversion factor is $\Delta_n=4$. The energy harvesting process follows a
stationary stochastic process. In addition, we run the simulation for $T=1000$
time slots in the Matlab software environment.

\begin{figure}[htbp]
\centering
\includegraphics[width=0.7%
\textwidth]{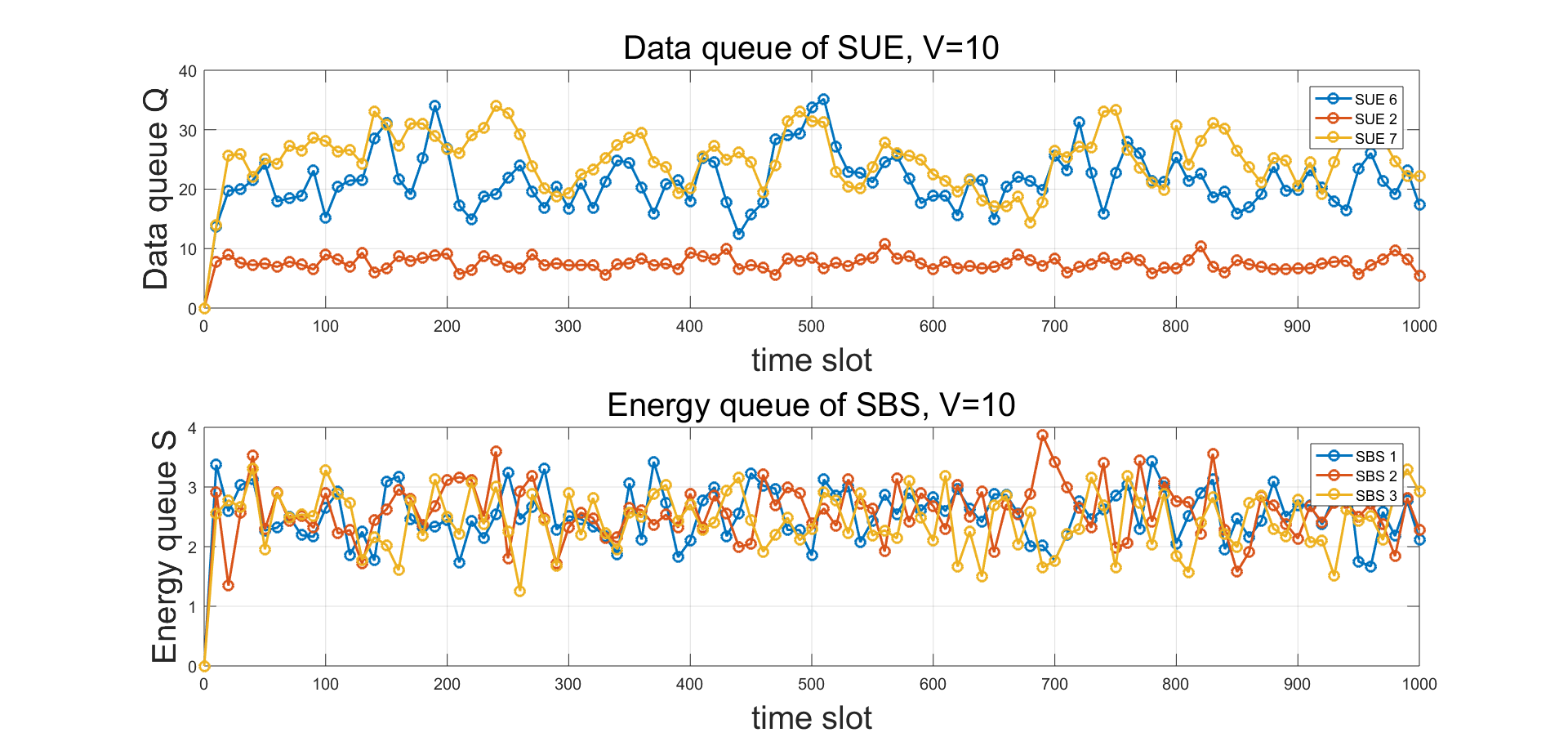}\newline
\caption{Dynamics of data queues and energy queues}\label{fig:dataener}
\end{figure}
\begin{figure}[htbp]
\centering
\includegraphics[width=0.7%
\textwidth]{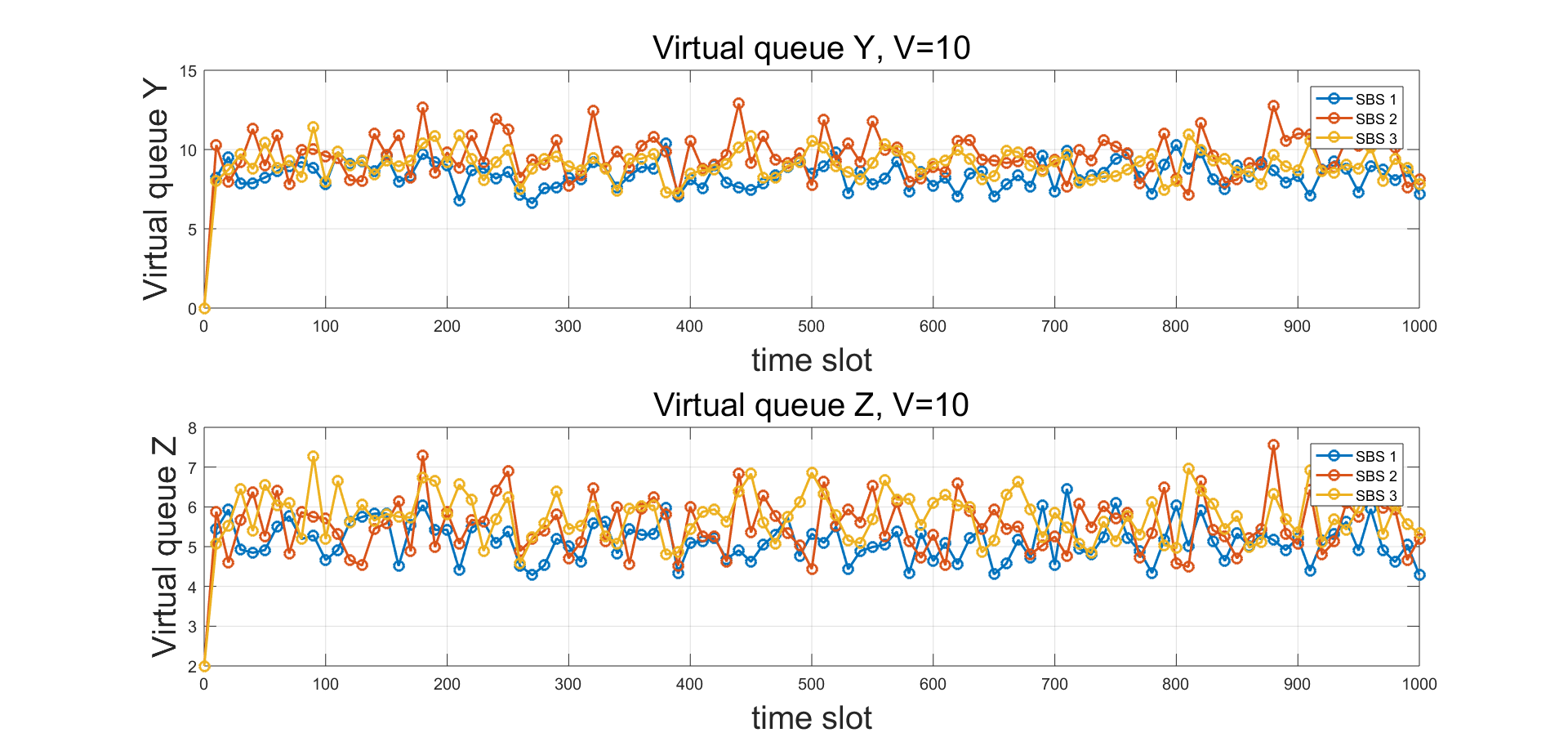}\newline
\caption{Dynamics of virtual queues Y and Z}\label{fig:queueyz}
\end{figure}

Firstly, we observe the queue stability in Fig. \ref{fig:dataener} and Fig. \ref{fig:queueyz} with $V = 10$.
Because all SUEs' data queues undergo similar trends, we take one SUE of
each SBS as example. Fig. \ref{fig:dataener} shows the dynamics of data queues $Q$ and energy queues $S$.
Fig. \ref{fig:queueyz} shows the dynamics of virtual queues $Y$ and $Z$. The upper figure in
Fig. \ref{fig:dataener} demonstrates the results in \textit{Theorem 2.a)} that the data
queues are bounded. The bottom figure in Fig. \ref{fig:dataener} shows the storage device is
below the capacity of storage device according to \textit{Theorem 1}. Moreover, in Fig. \ref{fig:queueyz},
we can also see that virtual queues $Y_n(t)$ and $Z_n(t)$ are bounded.
Hence, Fig. \ref{fig:dataener} and Fig. \ref{fig:queueyz} show that all the queues are bounded, which means that the network system is stabilized and that the long-term time-average constraints in optimization problem OP$_2$ are satisfied.

\begin{figure}[htbp]
\centering
\includegraphics[width=8cm]{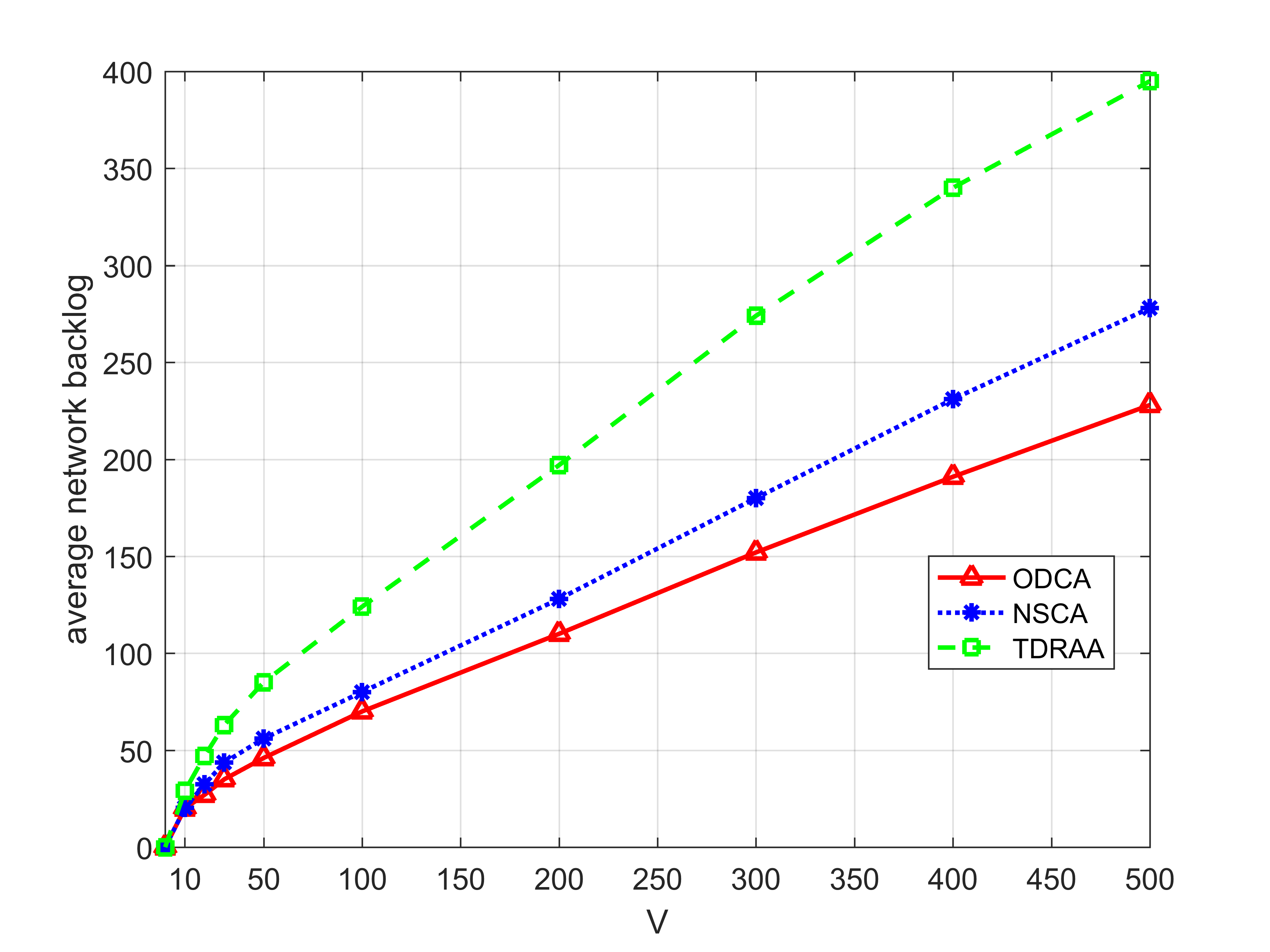}\newline
\caption{Average queue backlog versus parameter V}\label{fig:queuev}
\end{figure}
Fig. \ref{fig:queuev} shows the average network backlog, i.e., $\frac{1}{T U}%
\!\sum_{t=0}^{T-1}\!\sum_{n=1}^{N}\!\sum_{u=1}^{U_n}Q_{nu}(t)$, versus parameter $V$.
Compared with no spectrum sharing resource allocation algorithm
(NSRA) and time division multiple access resource allocation algorithm
(TDRAA), the average queue length in the proposed spectrum sharing algorithm is smaller than that in other
algorithms for any $V$, which suggests that the proposed spectrum sharing algorithm can reduce network
congestion.

\begin{figure}[htbp]
\centering
\includegraphics[width=8cm]{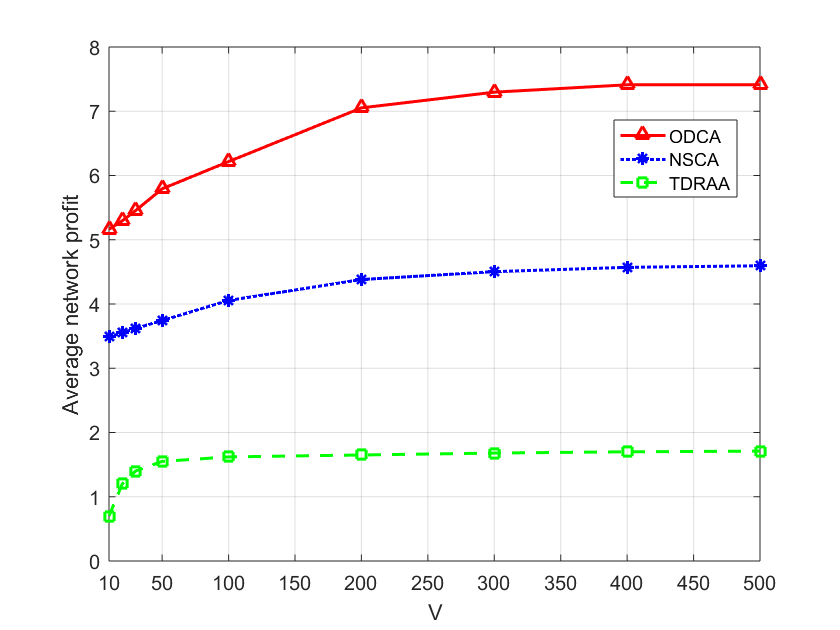}\newline
\caption{Performace comparision between proposed algorithm and other
algorithms}\label{fig:perfcom}
\end{figure}
Fig. \ref{fig:perfcom} shows the network performance comparison between the proposed
algorithm and the other two algorithms. Results show that the average network profits
obtained by sharing spectrum resources between SBSs are greater than
their selfish transmission. Furthermore, the performance of the proposed
algorithm is better than other algorithms, which demonstrates its effectiveness and feasibility.

\section{conclusion}
In this paper, we study the spectrum sharing problem between SBSs
powered by renewable energy. We allow the SBS to share free spectrum
resources and cooperate with other SBSs for dynamic resource allocation. A
multi-bargaining framework is modeled to measure the fairness of sharing profits of
SBSs. To solve the problem, we use Lyapunov optimization to decompose the
stochastic optimization problem. We develop an online dynamic control algorithm to
obtain the optimal transmission power and subchannel
assignment and the energy scheduling. Furthermore, the simulation results show the better performance
than other algorithms.
\begin{appendices}
\section{proof of lemma 1}
For any non-negative real numbers $x$, $y$ and $z$, there holds $\left ( \left | x-y \right |+z \right )^{2}\leq x^{2}+y^2+z^2-2x\left ( y-z \right )$. We can get the results as follow:
\begin{align}
Q_{nu}(t+1)^2\leq Q_{nu}(t)^{2}+R_{nu}(t)^{2}+D_{nu}(t)^{2}+2Q_{nu}(t)(D_{nu}(t)-R_{nu}(t))
\end{align}
\begin{align}
Y_{n}(t+1)^2\leq Y_{n}(t)^{2}+y_{n}^{in}(t)^2+y_{n}^{out}(t)^2+2Y_{n}(t)(y_{n}^{in}(t)-y_{n}^{out}(t))
\end{align}
\begin{align}
Z_{n}(t+1)^2\leq Z_{n}(t)^{2}+z_{n}^{in}(t)^2+z_{n}^{out}(t)^2+2Z_{n}(t)(z_{n}^{in}(t)-z_{n}^{out}(t))
\end{align}

Specially, the energy queue follows the inequality:
\begin{align}
(S_{n}&(t+1)-\rho_{n})^2-\left( S_{n}(t)-\rho_{n} \right)^2 \nonumber\\
&=\left ( S_{n}(t+1)+S_{n}(t)-2\rho_n \right )\left ( S_{n}(t+1)-S_{n}(t) \right )\nonumber\\
&=\left (2S_n(t)-2\rho_n+J_n(t)-F_n(t)  \right )\left ( J_n(t)-F_n(t) \right )\nonumber\\
&=2\left (S_n(t)-\rho_n \right )\left ( J_n(t)-F_n(t) \right )+\left ( J_n(t)-F_n(t) \right )^2 \nonumber\\
&\leq 2\left (S_n(t)-\rho_n \right )\left ( J_n(t)-F_n(t) \right )+J_n(t)^2+F_n(t)^2\nonumber\\
&\leq 2\left (S_n(t)-\rho_n \right )\left ( J_n(t)-F_n(t) \right )+J_n^{max2}+F_n^{max2}
\end{align}
where $J_n^{max}=\max\{S_n^{max},E_n^{max}\}$ and $F_n^{max}=S_n^{max}$.

By employing above inequalities, we can prove \textbf{Lemma 1}.
\section{proof of theorem 1}
We initialize the storage device $0\leq S_n(0)\leq S_n^{max}$ and suppose the limit holds for time slot $t$. We define $\rho_n=S_n^{max}-E_n^{max}$. Then we will prove it holds for the next time slot $t+1$. There are two cases to be considered:

(1) $0\leq S_n(t)\leq S_n^{max}-E_n^{max}$

It means $S_n(t)\leq \rho_n$, then
\begin{align}
J_n^*(t)=\min\{S_n^{max}-S_n(t),E_n(t)\}=E_n(t)
\end{align}

so the storage device level at the next time slot is
\begin{align}
S_n(t+1)&=\left[ S_n(t)-F_n^*(t)\right]^++J_{n}^*(t)\leq S_n(t)+E_n^{max}\leq S_n^{max}
\end{align}

(2) $S_n^{max}-E_n^{max}< S(t)\leq S_n^{max}$

It means $S_n(t)< \rho_n$. In this case, SBS will not accept renewable energy currently, i.e., $J_n^*(t)=0$ and then
\begin{align}
S_n(t+1)=S_n(t)-F_n^*(t)\leq S_n^{max}
\end{align}

In summary, $S_n(t+1)$ is bounded if $S_n(t)$ is bounded and this completes the proof.

\section{proof of theorem 2}
a) \textbf{Lemma 2.} For arbitrary small positive real number ${\varepsilon _1}$ and ${\varepsilon _2}$, there exists an
algorithm that makes independent, stationary and randomized
decisions  at each time slot based only on the observed network state, which satisfies
\begin{equation}
\mathbb{E}\left[ {D_{nu}^*\left( t \right) - R_{nu}^*\left( t \right)|\Theta \left( t \right)} \right] =  - {\varepsilon _1}
\end{equation}
\begin{equation}
\mathbb{E}\left[ {\mu _n^*\left( t \right) + \phi G_n^*\left( t \right) + C_n^{\min } - \sum\limits_{u = 1}^{{U_n}} {D_{nu}^*\left( t \right) - {O_n^*\left( t \right)}|\Theta \left( t \right)} } \right] =  - {\varepsilon _2}
\end{equation}
\begin{equation}
\mathbb{E}\left[ {J_n^*\left( t \right)|\Theta \left( t \right)} \right] = \mathbb{E}\left[ {F_n^*\left( t \right)|\Theta \left( t \right)} \right]
\end{equation}
where ${D_{nu}^*\left( t \right), {R_{nu}^*\left( t \right)}}, {\mu _n^*\left( t \right)}, {G_n^*\left( t \right)}, {O_n^*\left( t \right)}, {J_n^*\left( t \right)}, {F_n^*\left( t \right)}$ are corresponding results under the stationary algorithm. The similar proof of \textbf{Lemma 2.} can be found in \cite{31}

Since the algorithm is to minimize the right side of (34) under the constraints, we have
\begin{align}
& \Delta _V(t)\leq H+\sum\limits_{n=1}^{N}\sum\limits_{u=1}^{U_{n}}Q_{nu}(t)%
\mathbb{E}\left \{ D^*_{nu}(t)-R^*_{nu}(t)|\Theta (t) \right \}  \notag \\
&+ \sum\limits_{n=1}^{N}Y_{n}(t)\mathbb{E}\Bigg\{\mu^* _{n}(t)+C_{n}^{min} +\phi
G^*_{n}(t)-\sum_{u=1}^{U_{n}}{D^*}_{nu}(t)-O^*_n(t)|\Theta (t) \Bigg\}  \notag \\
&+ \sum\limits_{n=1}^{N}Z_{n}(t)\mathbb{E}\Bigg\{C_{n}^{min} +\phi
G^*_{n}(t)-\sum_{u=1}^{U_{n}}{D^*}_{nu}(t)-O^*_n(t)|\Theta (t) \Bigg\}  \notag \\
&+\sum\limits_{n=1}^{N}\Bigg\{\left( S_{n}(t)-\rho_n \right)\mathbb{E}\left
\{ J^*_{n}(t)-F^*_{n}(t)|\Theta (t) \right \}  \notag \\
&-V\sum_{n=1}^{N}\mathbb{E}\left \{ f(\mu^* _n(t))|\Theta (t) \right \}
\label{equ:lya}
\end{align}
where ${D_{nu}^*\left( t \right), {R_{nu}^*\left( t \right)}}, {\mu _n^*\left( t \right)}, {G_n^*\left( t \right)}, {O_n^*\left( t \right)}, {J_n^*\left( t \right)}, {F_n^*\left( t \right)}$ are corresponding results under the stationary algorithm referred to \textbf{Lemma 2}. Substituting (71) (72) and (73) into (74). we have
\begin{align}
 \Delta& _V(t)\leq  H- {\varepsilon _1}\sum\limits_{n = 1}^N {\sum\limits_{u = 1}^{{U_n}} {{Q_{nu}}\left( t \right)} } - {\varepsilon _2}\sum\limits_{n = 1}^N {{Y_n}\left( t \right)} - \left( {{\varepsilon _2} + {\mu^* _n}\left( t \right)} \right)\sum\limits_{n = 1}^N {{Z_n}\left( t \right)}- V\sum\limits_{n = 1}^N {f\left( {\mu _n^*\left( t \right)} \right)}
\end{align}

By taking iterated expectation and using telescoping sums over $t=0,1,...,T-1$, we get
\begin{align}\label{equ:cq}
&\mathbb{E}\left [ L(T)-L(0) \right ]-\sum\limits_{t = 0}^{T - 1} {V\mathbb{E}\left\{ {\sum\limits_{n = 1}^{N} {f\left( {\mu \left( t \right)} \right)} } \right\}}  \nonumber\\
&\leq TH-{\varepsilon _1}\sum\limits_{t = 0}^{T - 1} {\mathbb{E}\left[ {\sum\limits_{n = 1}^N {\sum\limits_{u = 1}^{{U_n}} {{Q_{nu}}\left( t \right)} } } \right]}  - {\varepsilon _2}\sum\limits_{t = 0}^{T - 1} {\mathbb{E}\left[ {\sum\limits_{n = 1}^N {{Y_n}\left( t \right)} } \right]}\nonumber\\
&  - \left( {{\varepsilon _2} + {\mu^* _n}\left( t \right)} \right)\sum\limits_{t = 0}^{T - 1} {\mathbb{E}\left[ {\sum\limits_{n = 1}^N {{Z_n}\left( t \right)} } \right] - V\sum\limits_{t = 0}^{T - 1} {\sum\limits_{n = 1}^N {f\left( {\mu _n^*\left( t \right)} \right)} } }
\end{align}

Considering $\mathbb{E}\left \{ L(T) \right \}>0$. we can find $\varepsilon {\rm{ = }}\min \left\{ {{\varepsilon _1},{\varepsilon _2}} \right\}$. Then we can get
\begin{align}
&\mathbb{E}\left [ L(T)-L(0) \right ]-\sum\limits_{t = 0}^{T - 1} {V\mathbb{E}\left\{ {\sum\limits_{n = 1}^{N} {f\left( {\mu \left( t \right)} \right)} } \right\}} \nonumber\\
&\leq TH- \varepsilon \sum\limits_{t = 0}^{T - 1} {\mathbb{E}\left\{ {\sum\limits_{n = 1}^N {\sum\limits_{u = 1}^{{U_n}} {{Q_{nu}}\left( t \right)} }  + \sum\limits_{n = 1}^N {{Y_n}\left( t \right) + \sum\limits_{n = 1}^N {{Z_n}\left( t \right)} } } \right\}}- V\sum\limits_{t = 0}^{T - 1} {\sum\limits_{n = 1}^N {f\left( {\mu _n^*\left( t \right)} \right)} }
\end{align}
and further get
\begin{align}
&\varepsilon \sum\limits_{t = 0}^{T - 1} {\mathbb{E}\left\{ {\sum\limits_{n = 1}^N {\sum\limits_{u = 1}^{{U_n}} {{Q_{nu}}\left( t \right)} }  + \sum\limits_{n = 1}^N {{Y_n}\left( t \right) + \sum\limits_{n = 1}^N {{Z_n}\left( t \right)} } } \right\}}  \nonumber\\
&\leq \mathbb{E}\left [ L(0) \right ]+TH+\sum\limits_{t = 0}^{T - 1} {VE\left\{ {\sum\limits_{n = 1}^{N} {f\left( {\mu \left( t \right)} \right)} } \right\}} - V\sum\limits_{t = 0}^{T - 1} {\sum\limits_{n = 1}^N {f\left( {\mu _n^*\left( t \right)} \right)} }
\end{align}

Dividing (77) by ${\varepsilon T}$ and considering $\mathbb{E}\left[ {L\left( 0 \right)} \right] = 0$, we have
\begin{align}
\underset{T\rightarrow \infty }{\mathrm{lim}}&\mathrm{sup}\frac{1}{T}\sum_{t=0}^{T-1}\mathbb{E}\left [ {\sum\limits_{n = 1}^N {\sum\limits_{u = 1}^{{U_n}} {{Q_{nu}}\left( t \right)} }  + \sum\limits_{n = 1}^N {{Y_n}\left( t \right) + \sum\limits_{n = 1}^N {{Z_n}\left( t \right)} } } \right ] \nonumber\\
&\leq \frac{{TH}}{{\varepsilon T}}+\frac{{V\left[ {\sum\limits_{t = 0}^{T - 1} {\mathbb{E}\left\{ {\sum\limits_{n = 1}^N {f\left( {{\mu _n}\left( t \right)} \right)} } \right\} - \sum\limits_{t = 0}^{T - 1} {\sum\limits_{n = 1}^N {f\left( {\mu _n^*\left( t \right)} \right)} } } } \right]}}{{\varepsilon T}}\nonumber\\
&\overset{\Delta }{=} \frac{{{H} + V\left( {\overline f  - {f^*}} \right)}}{{\varepsilon}}
\end{align}

Hence, the proof of \textit{Theorem 2.a)} is completed. Similarly, we can prove \textit{Theorem 2.b)}.
\end{appendices}

\end{document}